\documentclass [a4paper,12pt, twoside] {report}
 \usepackage{inputenc}
 \usepackage[english]{babel}
 \usepackage[dvips]{graphicx}
 \usepackage{epsfig, latexsym}

 \usepackage{amsfonts,amsmath,amssymb,amscd,array, euscript}

\begin {document}

 \newcommand {\bgc} {\begin{center}}
 \newcommand {\ec} {\end{center}}
\newcommand {\sms} {\smallskip}
 \newcommand {\mes} {\medskip}
 \newcommand {\nin}{\noindent}
 \newcommand {\etit}[1] {\bgc{\textbf{\large{#1}}} \mes\sms\ec}  
 \newcommand {\eauth}[2] {\bgc{\textbf{\large{#1}}\\ \mes\sms \emph{#2}} \mes\sms\ec}  
 \newcommand {\esect}[1] {\mes\mes\mes\sms\nin{\textbf{#1}} \mes\sms}  
 \newcommand{\gngfe}[2] {Hypercomplex Numbers in Geometry and Physics, \textbf{#1}, 200#2}

 \def\thebibliography#1{\subsubsection*{References}
 \list
  {[\arabic{enumi}]}{\settowidth\labelwidth{[#1]}\leftmargin\labelwidth
  \advance\leftmargin\labelsep
  \parsep=1pt plus 1pt
  \itemsep=\parsep
  \usecounter{enumi}}
  \def\newblock{\hskip .11em plus 0.33em minus -.07em}
  \sloppy
  \sfcode`\.=1000\relax}
 
 \catcode `\@=12
 \righthyphenmin=2
 \sloppy

\etit{THE NOTIONS OF DISTANCE\\ \mes AND VELOCITY MODULUS\\ \mes IN THE LINEAR
FINSLER SPACES}

\eauth{G.\,I. Garas'ko} {Electrotechnical Institute of Russia, Moscow\\
gigarasko@yandex.ru}

\eauth{D.\,G. Pavlov} {Bauman State Technical University, Moscow\\
hypercomplex@mail.ru}

{\small The formulas for the 3-dimensional distance and the velocity modulus in the
4-dimensional linear space with the Berwald-Moor metrics are obtained. The used
algorithm is applicable both for the Minkowski space and for the arbitrary
poly-linear Finsler space in which the time-like component could be chosen. The
constructed here modulus of the 3-dimensional velocity in the space with the
Berwald-Moor metrics coincides with the corresponding expression in the Galilean
space at small (non-relativistic) velocities, while at maximal velocities, i.e. for
the world lines lying on the surface of the cone of future, this modulus is equal to
unity. To construct the 3-dimensional distance, the notion of the surface of the
relative simultaneity is used which is analogous to the corresponding speculations
in special relativity. The formulas for the velocity transformation when one pass
from one inertial frame to another are obtained. In case when both velocities are
directed along one of the three selected straight lines, the obtained relations
coincide with the analogous relations of special relativity, but they differ in
other cases. Besides, the expressions for the transformations that play the same
role as Lorentz transformations in the Minkowski space are obtained. It was found
that if the three space coordinate axis are straight lines along which the
velocities are added as in special relativity, then taking the velocity of the new
inertial frame collinear to the one of these coordinate axis, one can see that the
transformation of this coordinate and time coordinate coincide with Lorentz
transformations, while the transformations of the two transversal coordinates differ
from the corresponding Lorentz transformations.}

\esect{Introduction}

The geometry of the classical (non-relativistic) space is usually connected with the
names of Galileo and Newton. It can be considered the second order approximation
with regard to the small parameter (the ratio of the velocity modulus to the speed
of light) of the Minkowski space geometry. But there are other geometries whose
metrics is not quadratic, for which the corresponding limit transition leads to the
Galilean space, that is, to the classical non-relativistic mechanics.

Starting with four dimensions that are definitely present in the physical world and
wishing first of all to regard the simplest metrics of the fourth order, it seems
necessary to begin with the linear space with the Berwald-Moor metrics. In one of
the basis its interval can be represented as the product of four coordinates
 \begin{equation}\label{1}
 S=\sqrt[4]{\xi_1\xi_2\xi_3\xi_4} .
 \end{equation}
This space we designate as  $H_4$ \cite{1}. The metrics function (\ref{1}) is a
particular case of the more general metrics function (\cite{2}, \cite{3})
 \begin{equation}\label{2}
 S=\xi_1^{(1+r_1+r_2+r_3)/4}\xi_2^{(1+r_1-r_2-r_3)/4}
 \xi_3^{(1-r_1+r_2-r_3)/4}\xi_4^{(1-r_1-r_2+r_3)/4} ,
 \end{equation}
for which all the parameters $r_1,\, r_2,\, r_3$ are set equal to zero. The
important property of $H_4$ is that it is connected with the commutative associative
algebra and has an analogue of the scalar product that can be introduced as a
symmetric poly-linear form of several vectors  \cite{4}.

Notice, that despite of its exotic view, the eq. (\ref{1}) metrics can be regarded
as a 4-dimensional generalization of the usual quadratic form characteristic for the
pseudo-Euclidean plane.
 \begin{equation}\label{3}
 S^2=x_0^2-x_1^2
 \end{equation}
in the special basis constructed out of the isotropic vectors can be presented as
 \begin{equation}\label{4}
 S^2=\xi_1\xi_2 .
 \end{equation}
This is already enough to expect the space with the eq. (\ref{1}) metrics to have
properties close to the properties of the pseudo-Euclidean space (especially,
2-dimensional one), one of such properties being certain relativistic features.

 \mes
\textit{Remark:} To simplify the formulas we will usually write the tensor indices
as subscripts and sometimes will not write the vector coordinates as the differences
between their ends. This should not lead to errors or misunderstandings since we use
only affine spaces and the lifting and lowering of the indices is not used.

\esect{Physical interpretation of the main geometrical objects}

Regarding a 4-dimensional multiple set as a model of the space-time, one should
first of all look for the effects taking place in its 3-dimensional subspace. The
last one should be preferably able to be interpreted as the regular 3-dimensional
classical space of the observer. In Minkowski space (and its Riemann
generalizations) the Euclidean properties of its 3-dimensional subspace are present
in the fundamental metrics form containing the positively defined quadratic
components. As a result of this, the methodological problems of comparing the
properties of such multiple sets with the properties of the real 3-dimensional space
(undoubtedly close to the Euclidean geometry), arise only as the corollaries of the
rejection of the absolute simultaneity.

Leaving the Minkowski space with its quadratic form for the Finsler space,
particularly to $H_4$, where the intervals are expressed by the fourth-order form,
the observer "living" in such a space could not be sure what kind of geometry he
finds around him. To answer this question let us find out which objects of this
multiple set are related to the common physical notions and values. But before that,
let us first give the interpretations of the analogues geometric objects connected
with the special relativity (SR). These interpretations are: \\
 1.  point in the 4-dimensional space -- event; \\
 2.  straight line -- world line of the inertial frame; \\
 3.  distance between the points on the straight line -- interval between the events;\\
 4.  set of isotropic (with the zero interval) straight lines crossing at one point -- light cone; \\
 5.  hyper-surface with the points that are equidistant from the fixed point --
 space-time hyper-sphere or set of events equidistant in observer's proper time from the fixed event; \\
 6.  hyper-surface with the points equidistant from the two fixed points --
 set of the relatively simultaneous events in the selected inertial frame
 whose world line passes through the fixed points; \\
 7.  straight lines parallel to the fixed line -- set of points that are motionless
in the 3-dimensional space of the observer located in the fixed inertial frame.

For geometrical objects in $H_4$ (as well as for many other Finsler spaces)
practically the same physical interpretations can be used. The differences reveal
themselves only in particular cases, and for $H_4$ they constitute the following
three facts: instead of a circular light cone there is a cone with flat sides; the
set of relatively simultaneous events (i.e. the set of events equidistant from the
two fixed points of space-time) is now not flat but it presents rather complicated
hyper-surface; and instead of the pseudo-Euclidean sphere consisting of three
hyperboloids (second order surfaces) there is now a hyper-surface consisting of 16
hyperboloids (fourth order surfaces). All these circumstances follow from the fact
that now the interval is not the square root of quadratic form, but the fourth order
root of the fourth order form (\ref{1}).

The special basis in which the $H_4$ interval has the laconic form (\ref{1}) is
connected with the special isotropic vectors. In the analogous basis in SR the
square of the interval looks rather unusually too:
 \begin{equation}\label{5}
 S^2=\xi_1\xi_2+\xi_1\xi_3+\xi_1\xi_4+\xi_2\xi_3
 +\xi_2\xi_4+\xi_3\xi_4 .
 \end{equation}

Such representation of the Minkowski space interval is rarely used, therefore, not
to step away from the usual SR constructions, let us transform the metrics of$H_4$
to the basis that is a Berwald-Moor analogue of the ortho-normal basis [5]. To do
this we use the linear substitution:
 \begin{equation}\label{6}
 \xi_i=A_{ij}x_j , \, \; (A_{ij})= \left(
 \begin{array}{rrrr}
  1 & 1 & 1 & 1 \\
  1 & 1 & -1 & -1 \\
  1 & -1 & 1 & -1 \\
  1 & -1& -1& 1
 \end{array} \right) ,  \, \;
 A_{ik}A_{kj}=4\delta_{ij}
 \end{equation}
-- and obtain the following expression for the fourth power of the interval :
 \begin{equation}\label{7}
 \begin{array}{r}
  S^4 = x_0^4-2x_0^2(x_1^2+x_2^2+x_3^2)+8x_0x_1x_2x_3+ \\
  +x_1^4+x_2^4+x_3^4 - 2x_1^2x_2^2-2x_1^2x_3^2-2x_2^2x_3^2 .
 \end{array}
 \end{equation}
Raising the square of the Minkowski interval
 \begin{equation}\label{8}
 S^2=x^2_0-x^2_1-x^2_2-x^2_3
 \end{equation}
to the second power to have the powers of four in both expressions, we get the
polynomial similar to that in the r.h.s of eq. (\ref{7}):
 \begin{equation}\label{9}
 S^4 = x_0^4-2x_0^2(x_1^2+x_2^2+x_3^2) + x_1^4+x_2^4+x_3^4 +
 2x_1^2x_2^2 + 2x_1^2x_3^2+2x_2^2x_3^2 .
 \end{equation}
In the regions characteristic for the non-relativistic physics where $|v_\alpha|\ll
1$, $v_\alpha=x_\alpha/x_0$, $\alpha=1, 2, 3$, the expressions (\ref{7}) and
(\ref{9}) coincide within the accuracy of the second power of small parameter
$|v_\alpha|$. This justifies the mentioned limit transition from the $H_4$ geometry
to the Galilean geometry, that is to the geometry of classical Newtonian physics.

 \clearpage
 \esect{Definitions of distance and velocity modulus in Minkowski space}

The observer in Minkowski space who attributes the equal distances (in the
3-dimensional sense) to a certain set of events can follow a simple geometric rule -
intercross two spheres with the same radii (hyperboloids in Minkowski space) located
in two different centers (Fig.1). The straight line passing through the centers of
these hyperbolic spheres can be associated with the inertial frame in which the
events on the cross-section of the corresponding hyperboloids appear to be
equidistant.

  \begin{figure}[!h]
  \centering
  \includegraphics[height=7cm]{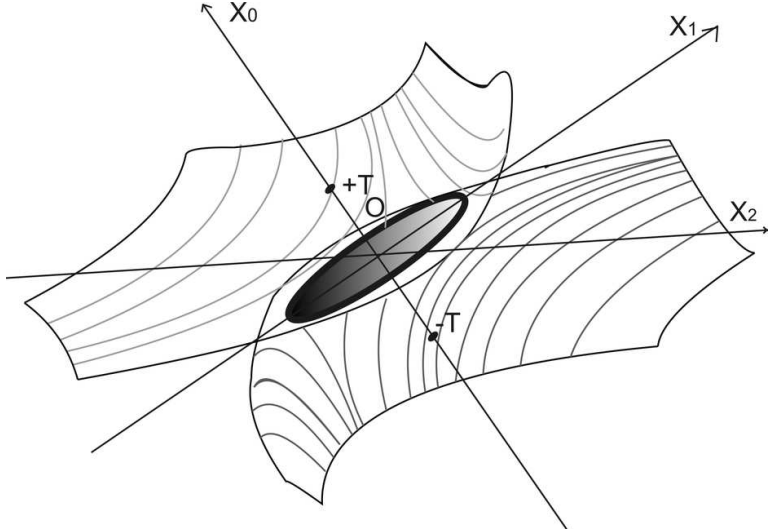}
  \caption{\small The cross-section of the hyperboloids in Minkowski space}
  \end{figure}

With no loss of generality, the centers of hyperboloids could be located on the time
axis $x_0$ symmetrically from the coordinates origin, i.e.,   in the points
$(-T,0,0,0)$ and $(T,0,0,0)$. To obtain the equation of the intersection of the two
pseudo Euclidean spheres with radii $S$ and with centers in these points, one has to
solve the system of equations
 \begin{equation}\label{10}
 \left.
 \begin{array}{l}
  S^2=(T+x_0)^2-x^2_1-x^2_2-x^2_3 , \\
  S^2=(T-x_0)^2-x^2_1-x^2_2-x^2_3 .
 \end{array}   \right\}
 \end{equation}
    After addition and subtraction of these equations, one gets
 \begin{equation}\label{11}
 \left.
 \begin{array}{l}
  S^2=T^2+x_0^2-x^2_1-x^2_2-x^2_3 , \\
 \; \; 0=2Tx_0 ,
 \end{array}   \right\}
 \end{equation}
The first of these equations describes the single-cavity spherical hyperboloid (time
axis is the axis of its symmetry), the second equation describes the hyper-plane
$x_0=0$ orthogonal to the $x_0$ axis (Fig.2).

  \begin{figure}[!h]
  \centering
  \includegraphics[height=7cm]{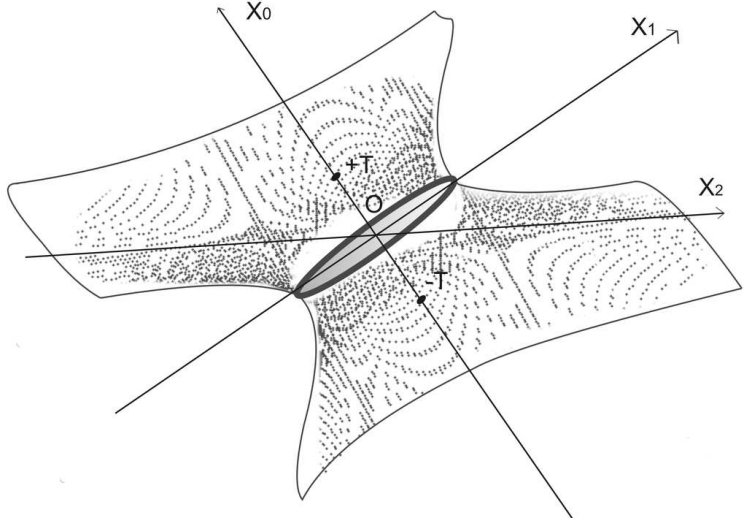}
  \caption{\small The cross-section of the plane and hyperboloid in Minkowski space}
  \end{figure}

Varying the value of the interval $S$ from $0$ to $T$, we get a set of 2-dimensional
surfaces put into one another. Each of these surfaces should be attributed a real
positive value, which the observer in the fixed frame could call 'distance'. To do
this, he should attribute these values to the arbitrary points of the surfaces and
expand these values on all the other points of the surfaces. The simplest
realization of this procedure is the drawing of a line across all the surfaces, then
the linear parameter along this line could be used as 'distance'. In particular, one
can use the $x_1$ axis as this line. Taking the linear rule for the correlation
between the distance $l$ and the coordinate $x_1$, i.e. substituting $x_0=0$,
$x_1=l$, $x_2=0$, $x_3=0$ into the first equation in (\ref{11}), we obtain the
expression
 \begin{equation}\label{12}
 l = \sqrt{T^2 - S^2} ,
 \end{equation}
Eq.(12) gives the relation between the distance $l$ and the radius (interval) $S$ of
the hyperboloids that were used to find the surfaces with the same value of
distances. This relation can be used to rewrite the first of eqs. (\ref{11}) in the
form of the well known in SR expression for the 3-dimensional distance between the
$x_0$ axis and the world lines parallel to it:
 \begin{equation}\label{13}
 l=\sqrt{x^2_1+x^2_2+x^2_3}  .
 \end{equation}

The described procedure of obtaining the expression for the distance is never used
in SR, but it is equivalent to one that is used. Here we need such a complicated
procedure to perform the analogous construction in $H_4$ space in which the SR
algorithms do not lead to the result.

To obtain the 3-dimensional velocity in Minkowski space one can use similar
speculations. Two points  $(x_{(1)0}$, $x_{(1)1}$, $x_{(1)2}$, $x_{(1)3})$ and
$(x_{(2)0}$, $x_{(2)1}$, $x_{(2)2}$, $x_{(2)3})$, (the second one is in the cone of
future of the first one) define a vector with coordinates $(x_{(2)i}-x_{(1)i})$ that
can be rewritten as
 \begin{equation}\label{14}
 (x_{(2)i}-x_{(1)i})\equiv (x_{(2)0}-x_{(1)0})v_i ,
 \end{equation}
where $v_0\equiv1$, and the components $v_1$, $v_2$, $v_3$  generate the
3-dimensional velocity vector. Then the interval for these three points can be
expressed with the help of the velocity components
 \begin{equation}\label{15}
 S_{21}=(x_{(2)0}-x_{(1)0}) \sqrt{1-v^2} ,
 \end{equation}
where
 \begin{equation}\label{16}
 v=\sqrt{v^2_1+v^2_2+v^2_3} .
 \end{equation}
Notice, that the modulus of the 3-dimensional velocity in SR has the property
 \begin{equation}\label{17}
 S_{21}=(x_{(2)0}-x_{(1)0})f(v) ,
 \end{equation}
where $f(v)$ is a function of one real variable. If vector $(1,v_1,v_2,v_3)$ and,
consequently, vector $(x_{(2)0}-x_{(1)0}$, $x_{(2)1}-x_{(1)1}$, $x_{(2)2}-x_{(1)2}$,
$x_{(2)3}-x_{(1)3})$ approach the isotropic direction, then $v\rightarrow 1$.

  \clearpage
 \esect{Definition of the distance and the velocity modulus in the $H_4$ space}

Let us take as a definition that in the $H_4$ space in the same way as in Minkowski
space, the cross-section of the two spheres (hyperboloids) with equal radii but
different centers is a set whose points are spatially equidistant from the observer
whose world line passes through these centers. Apart from the analogy with the
pseudo-Euclidean case, this statement is supported by the equality of the proper
times for the signals emitted from the $(-T,0,0,0)$ point and coming to the points
of the cross-section of two hyperboloids and the proper times of the back signals
emitted in these points and coming to the point $(T,0,0,0)$. From the point of view
of the observer whose world line passes through these points, i.e. coincides with
the $x_0$ axis, and who can use only the information concerning himself and these
signals, the latter reflect from the points of the 3-dimensional space equidistant
from the observer. The total travel time (on the "signals' watch") appears to be
equal to $2S$ for all pairs of signals and does not depend on the direction of
travel. The watch of the observer, who considers himself motionless, will read the
interval $2T$. Therefore, neither the readings of the signals' watch, nor the
readings of the observer's watch do not contradict the suggestion that the distances
from the observer to the world lines passing through the points of the cross-section
of the two hyperboloids, are the same. Consequently, they are completely
characterized by the two values $S$ and $T$.

To get the equation for the surface of the cross-section of the two hyperboloids
with the centers in the points $(-T,0,0,0)$ and $(T,0,0,0)$ in the $H_4$ space,
substitute first $(T+x_0,x_1,x_2,x_3)$ and then $(T-x_0,-x_1,-x_2,-x_3)$ instead of
$(x_0,x_1,x_2,x_3)$ into eq. (\ref{7}):
\begin{equation}\label{18}
\left.
\begin{array}{l}
    S^4 = (T+x_0)^4-2(T+x_0)^2(x_1^2+x_2^2+x_3^2)+8(T+x_0)x_1x_2x_3+ \\
  +x_1^4+x_2^4+x_3^4 - 2x_1^2x_2^2-2x_1^2x_3^2-2x_2^2x_3^2 , \\
    S^4 = (T-x_0)^4-2(T-x_0)^2(x_1^2+x_2^2+x_3^2)-8(T-x_0)x_1x_2x_3+ \\
  +x_1^4+x_2^4+x_3^4 - 2x_1^2x_2^2-2x_1^2x_3^2-2x_2^2x_3^2 .
\end{array}
\right\}
\end{equation}
Taking as in the case of Minkowski space the sum and the difference of these
equations, one gets
\begin{equation}\label{19}
\left.
\begin{array}{l}
  S^4=x_0^4 + 2x_0^2(3T^2-x_1^2-x_2^2-x_3^3)+8x_0x_1x_2x_3+ T^4 -\\
 -2T^2(x_1^2+x_2^2+x_3^2)+x_1^4+x_2^4+x_3^4-2(x_1^2x_2^2+x_1^2x_3^2+x_2^2x_3^2) ,  \\
\; \;  0=x_0^3+(T^2-x_1^2-x_2^2-x_3^2)x_0+2x_1x_2x_3  .
\end{array}\right\}
\end{equation}

  \begin{figure}[!h]
  \centering
  \includegraphics[height=7cm]{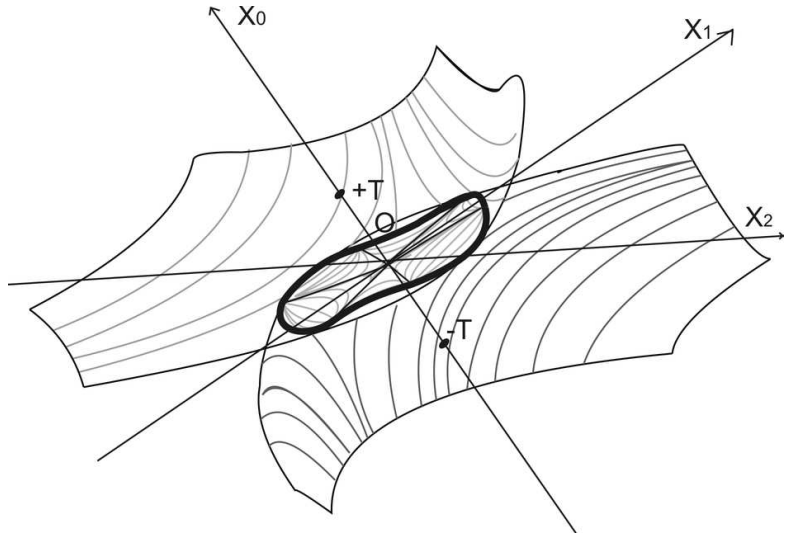}
  \caption{\small The cross-section of the hyperboloids in $H_3$ space}
  \end{figure}

It is not so easy to draw even schematically the 2-dimensional surfaces
corresponding to eq. (\ref{18}) in the 4-dimensional space. That is why to
illustrate the result we will use the similar surface in the 3-dimensional case,
Fig.3, corresponding to $H_3$ space, which is constructed similarly to $H_4$ and has
the following metrics in the isotropic basis
 \begin{equation}\label{20}
 S^3=\xi_1\xi_2\xi_3 .
 \end{equation}
Having passed from eqs. (\ref{18}) to eqs. (\ref{19}), we pass from the
cross-section of two hyperboloids to the cross-section of two new hyper-surfaces.
The first of them is in a sense equivalent to the single-cavern hyperboloid of the
Minkowski space, and the second is analogous to the hyper-plane $x_0=0$ of the
pseudo-Euclidean space, because there are equal intervals from every point of it to
the points $(-T,0,0,0)$ and $(T,0,0,0)$, Fig.4. But now the second equation of eqs.
(\ref{19}) defines the essentially nonlinear surface, this being the result of using
the Finsler metrics that has higher order than the quadratic one. From the physical
point of view, such hyper-surface could be related to the notion of relative
simultaneity. This is reasonable only in case when both the inertial frame is fixed,
and the characteristic scale $T$ (that gives the time between the instantaneous
location of the observer and the event with regard to which the simultaneity is
defined) are fixed. In the pseudo-Euclidean case this scale is unnecessary, since
the hyper-surface related to the notion of the relative simultaneity remained the
same for every interval separating the observer and the layer of the relatively
simultaneous events. In the linear Finsler spaces this is not so, and this leads to
the reconsideration of the properties of time, at least, for the spaces with the
non-quadratic metrics.

  \begin{figure}[!h]
  \centering
  \includegraphics[height=7cm]{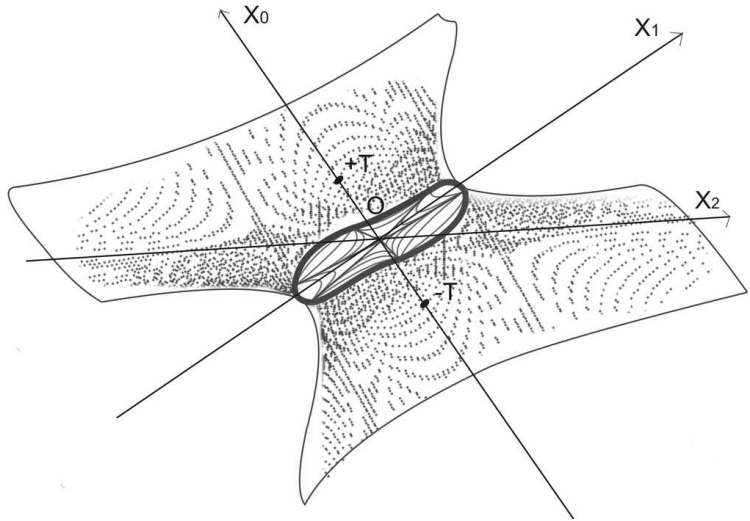}
  \caption{\small The cross-section of the special surface and hyperboloid in  $H_3$ space}
  \end{figure}

The cross-section of the hyperboloids (\ref{18}) with the centers at points $(-T,\;
0,\; 0,\; 0)$ and $(T$, $0$, $0$, $0)$ is such a set of events that the observer
whose world line passes through these points would consider equidistant from himself
(from his world line). Varying the interval $S$ from $0$ to $T$, we obtain the set
of 2-dimensional surfaces enclosed in each other, each of which corresponds to a
certain spatial distance. To characterize each of these 2-dimensional sets with one
and the same value of distance automatically, it is sufficient to attribute certain
values of distances to at least one of the points on each surface, and then extend
these values over all the points of the corresponding surface. As in the
pseudo-Euclidean case mentioned above, one can take a straight line crossing all
these surfaces, and call the linear parameter $l$ along this line the 'distance'
already not in the regular pseudo-Euclidean space, but in the linear Finsler
space-time.

The analysis of eq. (\ref{19}) shows that all the straight lines passing through the
coordinates origin and lying on one of the three planes $(x_1,x_2)$, $(x_1,x_3)$ or
$(x_2,x_3)$ belong to the surfaces of relative simultaneity of the $H_4$ space,
corresponding to this equation. Particularly, one of these lines is the $x_1$-axis,
therefore, relating the distance $l$ and the coordinate $x_1$, one gets the distance
$l$ from the observer to the motionless (with regard to him) observers for whom the
initial hyperboloids have the radii equal to $S$ and the half of the interval
between their centers is equal to $T$. Substituting $x_1=l, \; x_2=0, \; x_3=0$ into
eq. (\ref{19}), one gets
 \begin{equation}\label{21} \left.
\begin{array}{l}
  S^4=x_0^4 + 2x_0^2(3T^2-l^2) + T^4 - 2T^2l^2+l^4 ,  \\
\; \;  0=x_0^3+(T^2-x_1^2-x_2^2-x_3^2)x_0  .
\end{array}\right\}
\end{equation}
The second equation gives $x_0=0$, therefore, the first equation gives
 \begin{equation}\label{22}
  S^4 = T^4 - 2T^2l^2+l^4 .
 \end{equation}
Solving this equation for $l$, one gets
 \begin{equation}\label{23}
 l=\sqrt{T^2-S^2}  .
 \end{equation}
Thus, the 3-dimensional distance from the world line $(0,0,0)$ to the parallel world
line $(x_1,x_2,x_3)$ is expressed by the formula
 \begin{equation}\label{24}
 l(T,x_1,x_2,x_3)=\sqrt{T^2-S^2(T,x_1,x_2,x_3)} ,
 \end{equation}
where $S^2(T,x_1,x_2,x_3)$ is the square root of the r.h.s of the first of eqs.
(\ref{19}) in which $x_0$ is the real cubic root of the second of eqs. (\ref{19}))).

The expression for the 3-dimensional distance, being essentially different from the
regular spherically symmetric form (\ref{13}), contains the parameter $T$ lacking in
SR. These differences lead to rather unusual properties of the 3-dimensional
distances in $H_4$. Particularly, the distance from world line $A$A to the world
line $B$ is usually not equal to the distance from world line $B$ to the world line
$A$. But such effects reveal themselves only when any of the values $|x_\alpha|$ can
not be neglected with regard to $T$. If we can neglect the third and higher powers
of the ratio $|x_\alpha|/T$ with regard to unity, then the expression for the
distance (\ref{24}) takes the form
 \begin{equation}\label{25}
 l(T,x_1,x_2,x_3)\simeq \sqrt{x_1^2+x_2^2+x_3^2}  .
 \end{equation}

New qualitative feature that appears when constructing the surface of relatively
simultaneous events in $H_4$ and that distinguishes it from Minkowski space case is
the need for the concrete parameter $T$ measured in the units of length. It seems
logical to connect this characteristic scale, which is absent in SR, to the
observer, that is to the reference frame, and interpret it as an additional
parameter characterizing the reference frame and providing the possibility to
construct the fixed surface of relative simultaneity.

    Let us now pass to the 3-dimensional velocity.

Two points $(x_{(1)0},x_{(1)1}$, $x_{(1)2}$, $x_{(1)3})$ and $(x_{(2)0}$,
$x_{(2)1}$, $x_{(2)2}$, $x_{(2)3})$ in the $H_4$ space (the last point is in the
cone of future of the first one) define the vector with coordinates
$(x_{(2)i}-x_{(1)i})$ that can be rewritten with the help of velocity as
 \begin{equation}\label{26}
 (x_{(2)i}-x_{(1)i})\equiv (x_{(2)0}-x_{(1)0})v_i ,
 \end{equation}
where  $v_0\equiv1$, while the components $v_1$, $v_2$ and $v_3$ form the
3-dimensional velocity vector. Then the interval between these two points can be
expressed by the components of velocity as follows
 \begin{equation}\label{27}
 S_{21}=(x_{(2)0}-x_{(1)0}) \sqrt[4]{W} ,
 \end{equation}
where
 \begin{equation}\label{28}
 W=(1+v_1+v_2+v_3)(1+v_1-v_2-v_3)(1-v_1+v_2-v_3)(1-v_1-v_2+v_3) .
 \end{equation}
The modulus $v$ of the 3-dimensional velocity in $H_4$ must have the property
 \begin{equation}\label{29}
 S_{21}=(x_{(2)0}-x_{(1)0})f(v) ,
 \end{equation}
where $f(v)$ is a function of one real variable. If only one of the components of
the 3-dimensional velocity differs from zero, for example, $v_1$, then, naturally,
$v=|v_1|$, and expression (\ref{27}) gives
 \begin{equation}\label{30}
 S_{21}=(x_{(2)0}-x_{(1)0})\sqrt{1-v^2} .
 \end{equation}
In general case, the speculations similar to those for the 3-dimensional distance
give
 \begin{equation}\label{31}
 \sqrt{1-v^2}=\sqrt[4]{W(v_1,v_2,v_3)} ,
 \end{equation}
or
 \begin{equation}\label{32}
 v=\sqrt{1-\sqrt{W(v_1,v_2,v_3)}} .
 \end{equation}
In the non-relativistic approximation
 \begin{equation}\label{33}
 v\simeq \sqrt{v^2_1+v^2_2+v^2_3} .
 \end{equation}
If vector  $(1,v_1,v_2,v_3)$, and, consequently, vector   $(x_{(2)0}-x_{(1)0}$,
$x_{(2)1}-x_{(1)1}$, $x_{(2)2}-x_{(1)2}$, $x_{(2)3}-x_{(1)3})$ approach the
isotropic direction, then  $v\rightarrow 1$. Notice also, that in general case,
$W(-v_1,-v_2,-v_3) \neq W(v_1,v_2,v_3)$.

\esect{Addition of velocities}

The symmetry group $G_1(H_4)$ preserves invariant the interval (\ref{1}) and
consists of linear continuous transformations
 \begin{equation}\label{34}
 x'_i =\frac{1}{4} A_{ik} D_{km} A_{mj} x_j ,
 \end{equation}
where
 \begin{equation}\label{35}
 (D_{km})=diag(\exp{\varepsilon_0},\exp{\varepsilon_1},\exp{\varepsilon_2},\exp{\varepsilon_3}),
 \end{equation}
The real parameters $\varepsilon_i$ vary in $(-\infty,\infty)$ and suffice the
condition
 \begin{equation}\label{36}
 \varepsilon_0+\varepsilon_1+\varepsilon_2+\varepsilon_3=0,
 \end{equation}
This group can be parameterized with the three real values, $V_1$, $V_2$, $V_3$ that
can have the meaning of the components of velocity obtained by the motionless object
after the transformation ((\ref{35})
 \begin{equation}\label{37}
 \exp{\varepsilon_i}=\frac{A_{ij}V_j}{\sqrt{1-V^2}} ,
 \end{equation}
where $i,j=0,1,2,3$ ; $V_0=1$. If an object had the velocity components
$(v_1,v_2,v_3)$ in the initial reference frame, then in the new reference frame it
will have
 \begin{equation}\label{38}
 \left .
 \begin{array}{l}
 v'_1=\displaystyle\frac{v_1+V_1+v_2 V_3+v_3 V_2}{1+v_1V_1+v_2V_2+v_3V_3},  \\[15pt]
 v'_2=\displaystyle\frac{v_2+V_2+v_1 V_3+v_3 V_1}{1+v_1V_1+v_2V_2+v_3V_3},  \\[15pt]
 v'_3=\displaystyle\frac{v_3+V_3+v_1 V_2+v_2V_1}{1+v_1V_1+v_2V_2+v_3V_3} .
 \end{array} \right\}
 \end{equation}
The definition of the $G_1(H_4)$ group gives
 \begin{equation}\label{39}
 (x'_{(2)0}-x'_{(1)0})
 \sqrt{1-(v')^2}=(x_{(2)0}-x_{(1)0})\sqrt{1-v^2} ,
 \end{equation}
Thus, the formula for the 3-dimensional velocity in the new reference frame is
 \begin{equation}\label{40}
 v'=\sqrt{1-\frac{(1-v^2)(1-V^2)}{(1+v_1V_1+v_2V_2+v_3V_3)^2}} ,
 \end{equation}
because the transformations (\ref{34}) -- (\ref{37}) give
 \begin{equation}\label{41}
 x'_{(2)0}-x'_{(1)0}=\frac{1+v_1V_1+v_2V_2+v_3V_3}{\sqrt{1-V^2}}(x_{(2)0}-x_{(1)0}).
 \end{equation}

If the components of $v_\alpha$ and $V_\alpha$ contain only one component different
from zero each, and these correspond to the same specially chosen direction, for
example, $(v_1,0,0)$ and $(V_1,0,0)$, then the formulas (\ref{38}) coincide with the
corresponding formulas for addition of velocities in SR.

\esect{Transition from the motionless inertial frame to the moving one }

In this Section we will regard the transition from the old (no strokes) reference
frame to the new (stroked) inertial frame moving with the velocity $(V_1,V_2,V_3)$
relatively to the old one. That is, the point that has the velocity $(V_1,V_2,V_3)$
in the old frame will have the velocity $(0,0,0)$ in the new one. The formulas
(\ref{34}) -- (\ref{36}) will remain the same, while the formula (\ref{37}) will be
 \begin{equation}\label{42}
 \exp{(-\varepsilon_i)}=\frac{A_{ij}V_j}{\sqrt{1-V^2}} ,
 \end{equation}

That is, the transitions from one frame to another considered here and in the
previous Section are reverse to each other. Notice, that the change of
$(V_1,V_2,V_3)$ to $(-V_1,-V_2,-V_3)$ in (\ref{34}) -- (\ref{37}) does not give the
transition reverse to (\ref{34}) -- (\ref{37}).

So, the transition to the frame moving with velocity $(V_1,V_2,V_3)$ in the old
coordinates can be expressed by the new ones as
\begin{equation}\label{43}
\left(  \begin{array}{c}
  x_0 \\
  x_1 \\
  x_2 \\
  x_3
\end{array}   \right) = \frac{1}{4\sqrt{1-V^2}}\cdot \hat{A} \cdot
\left(
\begin{array}{c}
  (1+V_1+V_2+V_3)(x'_0+x'_1+x'_2+x'_3) \\
  (1+V_1-V_2-V_3)(x'_0+x'_1-x'_2-x'_3) \\
  (1-V_1+V_2-V_3)(x'_0-x'_1+x'_2-x'_3) \\
  (1-V_1-V_2+V_3)(x'_0-x'_1-x'_2+x'_3)
\end{array} \right) ,
\end{equation}
where matrix  $\hat{A}$ has the components $A_{ij}$ (\ref{6}).

Let us regard this transition for the case when all the components but one of the
velocity of the new frame in the old frame coordinates along the three special
directions are equal to zero, for example,  $V_1\neq 0$, but $V_2=0$ and $V_3=0$.
Then
\begin{equation}\label{44}
V=|V_1| ,
\end{equation}
and formulas (\ref{43}) take the form
\begin{equation}\label{45}
\left(
\begin{array}{c}
  x_0 \\
  x_1 \\
  x_2 \\
  x_3
\end{array} \right) =
\left(
\begin{array}{cccc}
  \frac{1}{\sqrt{1-V^2_1}} & \frac{V_1}{\sqrt{1-V^2_1}} & 0 & 0 \\
  \frac{V_1}{\sqrt{1-V^2_1}} & \frac{1}{\sqrt{1-V^2_1}} & 0 & 0 \\
  0 & 0 & \frac{1}{\sqrt{1-V^2_1}} & \frac{V_1}{\sqrt{1-V^2_1}} \\
  0 & 0 & \frac{V_1}{\sqrt{1-V^2_1}} & \frac{1}{\sqrt{1-V^2_1}}
\end{array} \right)  \cdot
\left(
\begin{array}{c}
  x'_0 \\
  x'_1 \\
  x'_2 \\
  x'_3
\end{array} \right)  ,
\end{equation}
or
\begin{equation}\label{46}
\left.
\begin{array}{cc}
  x_0=\displaystyle\frac{x'_0+V_1x'_1}{\sqrt{1-V^2_1}} & x_1=\displaystyle\frac{V_1x'_0+x'_1}{\sqrt{1-V^2_1}}  \\[15pt]
  x_2=\displaystyle\frac{x'_2+V_1x'_3}{\sqrt{1-V^2_1}} & x_3=\displaystyle\frac{V_1x'_2+x'_3}{\sqrt{1-V^2_1}}
\end{array}  \right\} .
\end{equation}

Such transformation of the coordinates $(x'_0,x'_1)\leftrightarrow (x_0,x_1)$
coincide with the corresponding transformation in SR, and the transformation
$(x'_2,x'_3)\leftrightarrow (x_2,x_3)$ differs from the corresponding transformation
in SR where $x_2=x'_2$, $x_3=x'_3$.

\esect{Conclusion}

The $H_4$ space which is the space of associative commutative hyper-complex numbers
(poly-numbers) is rather simple from the algebraic point of view - it is isomorphic
to the algebra of the square diagonal real matrices $4\times 4$. This space is an
anisotropic metric Finsler space with the three parametric Abel symmetry group and
it can not be reduced to a space with the quadratic metrics function. It is the
simultaneous consideration of the algebraic and geometric properties of $H_4$ that
leads to the appearance of a non-trivial mathematical object. As it was shown in
this paper, the consideration of the physical contents of $H_4$ together with its
algebraic and geometrical structures makes it even more complicated and interesting,
despite its initial algebraic simplicity: in the non-relativistic limit (neglecting
second and higher orders of the ratio of the velocity of the physical object to the
velocity of light), it is indistinguishable both from the Galilean space (the
classical mechanics space) and from the Minkowski space (SR). Moreover, even in the
general case there are some special directions and 2-dimensional planes for which
the properties of $H_4$ coincide with the corresponding properties of the Minkowski
space for the same directions and planes.

The difference between the $H_4$ space and the Minkowski space is due to the
anisotropy of the first one and to the physical effects proportional to the third
and higher powers of the ratio of the velocities of the physical objects to the
velocity of light. That is why, to our view, the question, which of the spaces is
most adequate for the description of the real World is open. In any case, the need
for the thorough investigation of the $H_4$ space and other similar spaces is even
more obvious, since they happened to be left aside from the mainstream of modern
geometric and physical research. This means that they could contain essences close
to the properties of the real World.

It should be underlined that the approach developed in this paper to get the modulus
of the 3-dimensional velocity and the 3-dimensional spatial distance is applicable
for any linear Finsler space for which there is a special coordinate system where
one time and three space coordinates can be separated.

\clearpage

\end{document}